\documentclass[showpacs, superscriptaddress,prd]{revtex4-1}

\usepackage{graphicx}
\usepackage{epsfig}
\usepackage{dcolumn}
\usepackage{bm}
\usepackage[utf8]{inputenc}
\inputencoding{utf8}
\usepackage{multirow}
\usepackage{caption}
\usepackage{subcaption}
\usepackage{amssymb}
\usepackage{amsmath}
\usepackage{hyperref}

\usepackage{amsmath}%
\newcommand\trick[1]{}

\def\be{\begin{equation}}
\def\ee{\end{equation}}

\def\ba{\begin{eqnarray}}
\def\ea{\end{eqnarray}}

\begin{document}

\title{Nonminimally Coupled Boltzmann Equation I: Foundations}

\author{Orfeu Bertolami}
\email{E-mail: orfeu.bertolami@fc.up.pt}
\affiliation{Departamento de F\'isica e Astronomia, Faculdade de Ci\^encias da Universidade do Porto, Rua do Campo Alegre s/n, 4169-007 Porto, Portugal}
\affiliation{Centro de Física do Porto, Rua do Campo Alegre s/n, 4169-007 Porto, Portugal}

\author{Cláudio Gomes}
\email{E-mail: claudio.gomes@fc.up.pt}
\affiliation{Centro de Física do Porto, Rua do Campo Alegre s/n, 4169-007 Porto, Portugal}

\date{\today}

\begin{abstract}
We derive the Boltzmann equation in the context of a gravity theory with non-minimal coupling between matter and curvature. We show that as the energy-momentum tensor is not conserved in these theories, it follows a condition on the normalisation of an homogeneous distribution function. The Boltzmann H-theorem is preserved such that the entropy vector flux is still a non-decreasing function in these theories. The case of an homogeneous and isotropic Universe is analysed.
\end{abstract}

\maketitle


\section{Introduction}
\label{sec:intro}

The Boltzmann equation is a microscopic statistical tool that describes the evolution of
the distribution function in phase space undergoing collisions. Mathematically, it corresponds to a partial integro-differential equation, whose exact solution is hard to find. Therefore, several approximations are commonly used to approach meaningful problems, for instance the relaxation time approximation and the Chapman-Enskog method. Despite its mathematical difficulties, the Boltzmann equation allows for the derivation of fundamental macroscopic equations, such as the Navier-Stokes equation for fluids and the Jeans and virial equations for self-gravitating systems \cite{binney}, the Maxwell-Vlasov equations for plasmas \cite{nicholson}, the Bloch-Boltzmann equations for electronic transport \cite{pottier}, and the relevant thermodynamic equation in an expanding Universe \cite{turner}. 

Notwithstanding, the Boltzmann equation has different formulations: the quantum, the classical, the relativistic and the general relativistic versions. One of its solutions is the Maxwell distribution, whose relativistic version has been derived in Ref. \cite{juttner1}. The relativistic Boltzmann equation may need to be extended in order to account for degenerate gases, having as solutions relativistic versions of the Bose-Einstein and the Fermi-Dirac distributions \cite{juttner2}.

Furthermore, its version for spacetime dynamics relies on General Relativity (GR), and allow for the description of, for instance, self-gravitating systems, primordial abundances and their evolution \cite{weinberg}. In fact, it is well known that GR is a well established gravity theory, in impressive agreement with Solar System and weak-field experiments \cite{gr0,gr1}. Nevertheless, it has raises questions both on theoretically and observationally: it is lacks a consistent quantum version, and at astrophysical and cosmological scales it requires the existence of two dark components to match observations \cite{whatif1,whatif2}. These two dark components together constitute around $95\% $ of the energy content of the Universe, and so far, neither of these components have been directly observed. As a consequence, several alternative theories of gravity have been put forward in the literature. The simplest generalisation of the Einstein’s theory are the so-called f(R) theories \cite{fr1,fr2,fr3}. But further generalisations are also admissible as is the case of a non-minimal matter-curvature coupling (NMC) \cite{nmc}. This model provides new insights on gravity and has a rich lore of implications for cosmology and astrophysics, as it mimics dark matter profiles at galaxies \cite{dm1} and clusters of galaxies \cite{dm2}, it accounts for the late time acceleration \cite{de1}, and is in agreement with data from inflation \cite{nmcinflation}, gravitational waves \cite{nmcgw} and the Abell 586 cluster virialisation \cite{linmc}.

Therefore, the aim of the present work is to generalise the Boltzmann equation in order to account for the non-minimal matter-curvature coupling model, and explore its main physical consequences. In this work, we shall use the (-+++) signature for the metric and work in units such that $c=1$ throughout the paper. Furthermore, we shall denote by $f(R)$ the general functions of the scalar curvature, whilst $f:=f(\vec{r},\vec{v},t)$ will represent the one-particle distribution function.

This work is organised as follows: in section \ref{sec:nmc}, one briefly introduces the non-minimal mater-curvature coupling model and its main features; in section \ref{sec:nmcboltzmann} we derive the Boltzmann equation in the context of this alternative gravity theory, in order to explore the conservation laws in section \ref{sec:laws}. We analyse the case of an homogeneous and isotropic universe in Sec. \ref{sec:universe}. We present our conclusions in Sec. \ref{sec:conclusions}.


\section{The non-minimal matter-curvature coupling model}
\label{sec:nmc}

In the extended $f(R)$ theories with a non-minimal coupling between curvature and matter, the action functional reads \cite{nmc}:
\begin{equation}
\label{modelo}
S=\int d^4x \sqrt{-g} \left[\kappa f_1\left(R\right) + f_2 \left(R\right)\mathcal{L}\right]~,
\end{equation}
where $f_1(R), f_2(R)$ are arbitrary functions of the Ricci scalar $R$, $\kappa=c^4/(16\pi G)$ and $\mathcal{L}$ is the matter Lagrangian density.

Varying the action with respect to the metric, $g_ {\mu\nu}$, leads to the following field equations:
\begin{equation}
\label{fieldequations}
\left( F_1(R) + \frac{F_2(R)\mathcal{L}}{\kappa}\right) G_{\mu\nu} = \frac{1}{2\kappa}f_2(R) T_{\mu\nu} + \Delta_{\mu\nu}\left( F_1(R) + \frac{F_2(R)\mathcal{L}}{\kappa}\right) + \frac{1}{2}g_{\mu\nu}\left(f_1(R)-F_1(R)R-\frac{F_2(R)R\mathcal{L}}{\kappa}\right) ~,
\end{equation}
where $G_{\mu\nu}:=R_{\mu\nu}-(1/2) g_{\mu\nu}R$ is the Einstein tensor, $F_i(R) \equiv df_i(R)/dR$ and $\Delta_{\mu\nu}\equiv \nabla_{\mu}\nabla_{\nu} - g_{\mu\nu} \square $. It is straightforward to see that choosing $f_1(R)=R$ and $f_2(R)=1$, one recovers General Relativity.

Taking the trace of the previous equations, one gets:
\begin{equation}
\label{trace}
\left(F_1(R)+\frac{F_2(R)\mathcal{L}}{\kappa}\right)R-2f_1(R)=-3\square \left(F_1(R)+\frac{F_2(R)\mathcal{L}}{\kappa}\right) + \frac{1}{2\kappa}f_2(R)T ~.
\end{equation}

Using the Bianchi identities in the field equations, Eq. (\ref{fieldequations}), one finds that the energy-momentum tensor is no longer (covariantly) conserved in this model:
\begin{equation}
\label{nonconservation}
\nabla_{\mu}T^{\mu\nu} = \frac{F_2(R)}{f_2(R)} \left(g^{\mu\nu}\mathcal{L}-T^{\mu\nu}\right) \nabla_{\mu}R ~.
\end{equation}

This feature implies that for a perfect fluid, test particles do not follow geodesics. In fact, the geodesics equation in these theories read \cite{nmc}:
\begin{equation}
\frac{du^{\alpha}}{ds}+\Gamma^{\alpha}_{\mu\nu}u^{\mu}u^{\nu}=\mathfrak{f}^{\alpha} ~,
\end{equation}
where the extra force, per unit mass, for a perfect fluid is given by:
\begin{equation}
\mathfrak{f}^{\alpha}=\frac{1}{\rho+p}\left[\frac{F_2(R)}{f_2(R)}\left(\mathcal{L}_m-p\right)\nabla_{\nu}R-\nabla_{\nu}p\right]h^{\alpha\nu}~,
\end{equation}
where $h^{\alpha\nu} = g^{\alpha\nu} + u^{\alpha}u^{\nu}$ is the projection operator and $u^\mu$ denotes the particle's 4-velocity. In these theories, the degeneracy between the Lagrangian choices $\mathcal{L}=-\rho$ or $\mathcal{L}=p$ for perfect fluids is lifted, in opposition to what happens in GR \cite{lagrangian choices GR}, since it yields different behaviours for the extra force term (See Ref. \cite{lagrangian choices} for a thorough discussion).

For a matter Lagrangian of the form $\mathcal{L}=-\rho$ and $\nabla_{\nu}p=0$, the extra force reads:
\begin{equation}
\mathfrak{f}^{\alpha}=-\partial_{\nu}\Phi_c h^{\alpha\nu}~.
\end{equation}
where one has defined the non-minimal coupling potential as $\Phi_c := \ln (f_2(R))$ \cite{martins}.
%

\section{The Boltzmann equation in the NMC model}
\label{sec:nmcboltzmann}

In statistical physics, the Liouville theorem which states that the phase-space distribution function, $f$, is constant along the trajectories of the system when no collisions occur, which implies that the distribution function behaves as an incompressible fluid. However, when particles collide, we have that the rate of change in phase-space is given by:
\begin{equation}
\frac{df}{dt}=\left(\frac{df}{dt}\right)_{coll.} ~.
\end{equation}

From this equation, one can derive the so-called Boltzmann equation, which in its general relativistic version reads \cite{stewart}:
\begin{equation}
p^{\mu}\frac{\partial f}{\partial x^{\mu}}+\frac{d p^{\mu}}{d \tau^*}\frac{\partial f}{\partial p^{\mu}}=\left(\frac{\partial f}{\partial \tau^*}\right)_{coll.}~,
\end{equation}
where $f=f(x^{\mu},p^{\mu})$ is the distribution function which depends in the spacetime coordinates, $x^{\mu}$, and in the 4-momentum, $p^{\mu}$, and $\tau^* := t/m$ is the affine parameter.

Given the extra force in the NMC theories, the Boltzmann equation is modified to:
\begin{equation}
p^{\mu}\frac{\partial f}{\partial x^{\mu}}-\left(\Gamma^{\sigma}_{\mu\nu}p^{\mu}p^{\nu}-m^2\mathfrak{f}^{\sigma}\right)\frac{\partial f}{\partial p^{\sigma}}=\left(\frac{\partial f}{\partial \tau^*}\right)_{coll.}~.
\end{equation}

In fact, the non-minimal coupling model introduces a term related to the extra force in the Boltzmann equation, which depends on the matter Lagrangian density. We can further consider the mass-shell conditions $p_0=\sqrt{g_{00}m^2+(g_{0i}g_{0j}-g_{00})g_{ij}p^ip^j}$, $p^0=(p_0-g_{0i}p^i)g^{00}$, i.e., $p^{\mu}p_{\mu}=-m^2$. Taking into consideration the previous relation, one gets:

\begin{equation}
p^{\mu}\frac{\partial f}{\partial x^{\mu}}-\left(\Gamma^{i}_{\mu\nu}p^{\mu}p^{\nu}-m^2\mathfrak{f}^{i}\right)\frac{\partial f}{\partial p^{i}}=\left(\frac{\partial f}{\partial \tau^*}\right)_{coll.}~.
\end{equation}

In fact, we call the Lioville operator to the operator acting on distribution function on the left hand side of the previous equation, i.e. $L[f]:=df/dt =\left[p^{\mu}\frac{\partial }{\partial x^{\mu}}-\left(\Gamma^{i}_{\mu\nu}p^{\mu}p^{\nu}-m^2\mathfrak{f}^{i}\right)\frac{\partial }{\partial p^{i}}\right] f$. The right hand side (RHS) of the previous equation may, in general, be very hard to compute. We shall, for simplicity, consider only binary collisions. This is a well motivated approximation since we are studying collisions in very short periods of time, during which one-to-one collisions are the dominant type. Furthermore, these are usually accompanied by the following assumptions:
\begin{enumerate}
\item[i)] The fluid under considerations is diluted enough;
\item[ii)] The particles (molecules) have no internal structure, ensuring the elasticity of the collisions;
\item[iii)] The external forces do not affect the collisions, thus collisions are local and instantaneous;
\item[iv)] Any previous correlations between velocities of the particles prior to the collision are neglected, an assumption known as the molecular chaos hypothesis (\textit{Stosszahlansatz}), which ensures the irreversibility of the kinetic equation.
\end{enumerate}

Bearing this in mind, the collisional term of the RHS of the Boltzmann equation can be written, in general, as \cite{mathematical}
\begin{equation}
\left(\frac{\partial f}{\partial t}\right)_{coll.}=\int W(p' + p'_2 \to p + p_2)\left(f'f'_2-f f_2\right)\pi(p')\pi(p'_2)\pi(p_2)~,
\end{equation}
where $\pi(p_*)$ is the invariant volume in the momentum space, and $W(p' + p'_2 \to p + p_2)$ is the transition probability, which obeys the following relations:
\begin{eqnarray}
&&W(p' + p'_2 \to p + p_2)=W(p'_2 + p' \to p_2 + p)\nonumber\\
&&W(p' + p'_2 \to p + p_2)=W(p_2 + p \to p'_2 + p')~.
\label{symmetries}
\end{eqnarray}

The first symmetry is trivial, whist the second exhibits the principle of detailed balance, that is, the microscopic reversibility \cite{israel}.

In particular, we can introduce the differential cross-section, $\sigma$ , and the invariant flux, $F=\sqrt{\left((p_2^{\mu}p_{\mu})^2-m^4\right)}$, of the collisions in such a way that the collision integral can be rewritten as:

\begin{equation}
\left(\frac{\partial f}{\partial t}\right)_{coll.}=\int\left(f'_2f'-f_2f\right)F\sigma d\Omega\sqrt{-g}\frac{d^3p_2}{(p_{2})_{0}}~,
\end{equation}
where $d\Omega$ the infinitesimal solid angle for the binary collisions.

\section{Conservation laws}
\label{sec:laws}

Let $\Psi(x,p)$ be a smooth function defined on the phase space for each particle such that for binary collisions:
\begin{equation}
\Psi(x,p'_2)+\Psi(x,p')-\Psi(x,p)-\Psi(x,p_2) = 0~.
\end{equation}

Such a function is called an collisional invariant. It can be shown that the most general collisional invariant is of the form \cite{mathematical, kremer, book1}:
\begin{equation}
\Psi(x,p)=\beta_{\mu}(x) p^{\mu}+ \alpha ~,
\end{equation}
where $\alpha$ is a linear combination of scalar quantities conserved in the collisions.

Taking this into consideration, let us multiply both sides of the Boltzmann equation by $\Psi$ and integrate over the volume element $\pi(p)$. We are left with
\begin{eqnarray}
\int \Psi(x,p)L[f]\pi(p)= -\frac{1}{4}\int W(p'+p'_2 \to p+p_2) \left[\Psi(x,p')+\Psi(x,p'_2)-\Psi(x,p)-\Psi(x,p_2)\right]\times \nonumber\\ 
\times \left[f'f'_2-ff_2\right]\pi(p'_2)\pi(p')\pi(p_2)\pi(p)~,
\label{eqn:transfer}
\end{eqnarray}
where we have used the symmetries of the transition probability scalar, Eq. (\ref{symmetries}).

 It is straightforward to check that collisional invariants leave the RHS of the Boltzmann equation identically null.
 
Some important physical quantities can be written as moments of the distribution function. In particular, we refer:
\begin{eqnarray}
&&A := \int f \pi ~, \\
&&N^{\mu} := \int p^{\mu} f \pi ~,	\\
&&T^{\mu\nu} :=\int p^{\mu}p^{\nu} f \pi ~,\\
\end{eqnarray}
where the first quantity is a real valued function of the position which is equal to 1 if we require a normalised homogeneous distribution function, the second corresponds to the flux density of particles which crosses a given 3-surface element, and the last one matches the definition of the energy-momentum tensor.

In fact, one can take the covariant derivative of the previous quantities, finding that the first trivially vanishes. For the second and third quantities, it can be shown that they lead to (likewise to the cases of Refs.\cite{stewart,mathematical}):
\begin{eqnarray}
&&\nabla_{\mu}N^{\mu} = \int L[f] \pi ~,\label{eqn:particleflux}\\
&&\nabla_{\mu}T^{\mu\nu}=\int L[f]p^{\nu}\pi + m^2\mathfrak{f}^{\nu}A ~.
\end{eqnarray}

Since the first equality above corresponds to the collisional invariant scalar $\Psi=1$, hence the RHS of Eq.(\ref{eqn:transfer}) vanishes. It follows that $\nabla_{\mu}N^{\mu}=0$ and the flux density of particles is covariantly conserved. As for the divergence of the energy-momentum tensor, and from the identification $\Psi\to\Psi^{\mu}=p^{\mu}$ into Eq.(\ref{eqn:transfer}), we get:
\begin{equation}
\nabla_{\mu}T^{\mu\nu}= m^2\mathfrak{f}^{\nu}A ~.
\end{equation}

Expanding the previous equation, and noting that $\nabla_{\nu}T^{\mu\nu}=(\rho+p)\mathfrak{f}^{\mu}+\nabla_{\nu}p ~h^{\mu\nu}$, we obtain:
\begin{equation}
A=\frac{\rho + p}{m^2}\frac{(\mathcal{L}-p)\nabla_{\nu}\ln f_2(R)}{(\mathcal{L}-p)\nabla_{\nu}\ln f_2(R) - \nabla_{\nu} p}~,
\end{equation}
which in the non-relativistic limit gives $A \sim \rho /m^2$. We should note that $A$ is not just a normalisation function as it is defined as the integral of the distribution function over the momentum space and not over the whole phase space.

A physical quantity of great importance in kinetic theory is the entropy flux vector field \cite{mathematical}:
\begin{equation}
S^{\mu} := -\int f \log (\xi f) p^{\mu} \pi ~,
\end{equation}
where $\xi=h^3/r$, with $h$ being the Planck constant and $r$ the spin degeneracy of the each particle \cite{stewart}. By taking the divergence of the entropy flux vector and by the same token as in Eq. (\ref{eqn:particleflux}), we get:
\begin{equation}
\nabla_{\mu}S^{\mu}=-\int L[f \log (\xi f)] p^{\mu} \pi ~,
\end{equation}
which can be rewritten as:
\begin{equation}
\nabla_{\mu}S^{\mu}=-\int \left(1+\log (\xi f) \right) L[f] p^{\mu} \pi ~,
\end{equation}
which in its turn can be compared with Eq. (\ref{eqn:transfer}), resulting in the identification $\Psi=1+\log (\xi f)$. It is then straightforward to get:
\begin{eqnarray}
\nabla_{\mu}S^{\mu}=\frac{1}{4}\int W(p+p_2 \to p'+p'_2) \left[\log (f f_2)-\log (f' f'_2)\right]\times \nonumber\\ 
\times \left[ff_2-f'f'_2\right]\pi(p)\pi(p_2)\pi(p')\pi(p'_2)~.
\end{eqnarray}

Since $W(p+p_2 \to p'+p'_2)$ is positive, the distribution functions are assumed to be strictly positive, together with the inequality
\begin{equation}
(\log y- \log x)(y-x)\geq 0~,
\end{equation}
which holds for $\forall x,y > 0$, we get that 
\begin{equation}
\nabla_{\mu}S^{\mu}\geq 0 ~.
\end{equation}

This is a quite relevant result as it expresses that entropy of a system in the presence of a non-minimal coupling is a non-decreasing vector function, since in these theories the Liouville operator has only an extra force term that does not change through collisional processes. In other words, the Boltzmann H-theorem is preserved in these theories as in General Relativity.

We should however note that there is also the Gibbs H-theorem which relies on the full distribution function in contrast with the one-particle distribution function of the Boltzmann H-theorem \cite{gibbsboltzmann}. They differ by very small corrections unless mixtures of different particle species are present, which is not the present case of this work. Therefore, we expect they both hold for this alternative gravity model.

Given these results for the nonminimally coupled Boltzmann equation, it is now relevant to study some cosmological implications, such as for a homogeneous and isotropic Universe.

\section{Homogeneous and isotropic Universe}
\label{sec:universe}

In an isotropic and homogeneous Universe, the metric is the Robertson-Walker one:
\begin{equation}
ds^2=-dt^2+a^2(t)d\vec{x}^2~, 
\end{equation}
where $a(t)$ is the scale factor.

Therefore, the time-time component of the field equations, Eq. \ref{fieldequations}, for a perfect fluid, gives a modified Friedmann equation \cite{modified friedmann}:
\begin{equation}
6H^2=\frac{1}{\Theta}\left[\frac{f_2(R)\rho}{\kappa}-6H\partial_t\Theta+\Theta R- f_1(R)\right]~,
\label{eqn:modifiedfriedmann}
\end{equation}
where $\Theta:= F_1(R)+\frac{1}{\kappa}F_2(R)\mathcal{L}$, $H=\dot{a}/a$ is the Hubble parameter, and the space-space components lead to a Raychaudhury equation \cite{modified friedmann}:
\begin{equation}
-\Theta\left(2\dot{H}+3H^2\right)=f_2(R)p+\frac{1}{2}\left(f_1(R)-\Theta R\right)~.
\end{equation}

As for the Boltzmann equation, we note that the extra-force term vanishes $f^i \sim -\partial_{\nu}\Phi_ch^{i\nu}=-a^2\partial_i\Phi_c \sim \partial_i R = 0$ in the comoving frame, $u^{\mu}=(1,0,0,0)$. Thus, and taking into account the on-shell mass condition, $p^{\mu}p_{\mu}=-m^2$, the Boltzmann equation reads:
\begin{equation}
p^{0}\frac{\partial f}{\partial x^{0}}-2Hp^0p^i\frac{\partial f}{\partial p^{i}}=\int\left(f'_2f'-f_2f\right)F\sigma d\Omega\sqrt{-g}\frac{d^3p_2}{(p_{2})_{0}}~.
\end{equation}

This is formally identical to the Boltzmann equation for a homogeneous and isotropic Universe in General Relativity, except by the fact that the scale factor evolution is now given by the modified Friedmann equation, Eq. (\ref{eqn:modifiedfriedmann}).

In order to solve the previous equation for the distribution function, we need to simplify its right hand side. One way to do that is to assume the Anderson and Witting model together with the relaxation time expansion \cite{anderson, kremer}:
\begin{equation}
C[f]\equiv \left(\frac{\partial f}{\partial t}\right)_{coll.} =-\frac{1}{\tau}\left(f-f^{(0)}\right)~, 
\end{equation}
where $\tau$ is the characteristic time which is of order of the mean free time and $f^{(0)}$ is the equilibrium  Maxwell-Jüttner distribution function:
\begin{equation}
f^{(0)}:= \frac{n}{4\pi m^2 k_B T K_2(m/k_BT)} \exp \left[-\frac{\sqrt{m^2+a^2|\vec{p}|}}{k_B T}\right]~,
\end{equation}
given in terms of the temperature $T$, particle number density $n$, Boltzmann constant $k_B$ and the modified Bessel function of the second kind
$K_n(x)=\int_0^{+\infty}e^{-x \cosh y}\cosh (ny) ~dy$.

This Anderson-Witting model is a generalisation of the relativistic Marle \cite{marle1,marle2,marle3} and the non-relativistic BGK (Bhatnagar, Gross and Krook) \cite{BGK} models.

Using the Chapman-Enskog  method, we search for a solution of this Boltzmann equation of the form:
\begin{equation}
f=f^{(0)}(1+\phi) ~,
\end{equation}
where $\phi$ measures the deviation from equilibrium. Thus, we obtain \cite{kremer}:
\begin{equation}
f=\sum_{l=0}^{+\infty}\left[-\tau\left(\frac{\partial}{\partial x^0}-2Hp^i\frac{\partial}{\partial p^i}\right)\right]^l f^{(0)}~.
\end{equation}
However, this series does not converge in general, unless the indexed expression has a norm less than unity. Considering the asymptotic $\tau\to 0$, we can find \cite{kremer}:
\begin{equation}
f\approx f^{(0)} \left\lbrace 1-\tau \left[  \frac{\dot{n}}{n}+\left(1-\zeta\frac{K_3}{K_2}+\frac{1}{k_BT}p_0\right)\frac{\dot{T}}{T}+\frac{1}{k_BT}H\frac{|\vec{p}|^2}{p_0}\right] \right\rbrace~,
\end{equation}
where $\zeta := m/k_B T$. Although this solution could be seen as formally equivalent to General Relativity, this is not completely true. In fact, as in GR, the particle number density is conserved, hence $\dot{n}+3Hn=0$. However, the time evolution of the temperature is not trivial as it relies on the non-conservation of the energy-momentum tensor and explicitly depends on the matter Lagrangian choice and on the form of the non-minimal coupling function, $f_2(R)$. In fact, from the fundamental thermodynamic relation we obtain: $T=\left(\frac{\partial U}{\partial S}\right)_{V,N}$, i.e., temperature is expressed as the  the rate of change of internal energy with respect to entropy, provided that both volume, $V$ and number of particles, $N$, are held constant. Hence, we need to carefully study the thermodynamic processes which a perfect fluid could undergo. We point out, for instance, that it was shown from gravitational baryogenesis arguments that the relative time change of entropy is as $\dot{S}/S \sim (F_2(R)/f_2(R)) \dot{R}$ \cite{baryogenesis}. Furthermore, we need to express the equation of state parameter, $w=p/\rho$, which renders different solutions. In fact, in an expanding Universe, relativistic matter does not feel the non-minimal coupling effect, hence $\rho_{rel}\sim a^{-4}$, however, radiation has the behaviour $\rho_r \sim a^{-4}f_2(R)^{-4/3}$ \cite{reheating}. Also non-relativistic matter fields evolve differently than their counterpart in GR \cite{reheating}. Therefore, we can have very different solutions for the distribution function in theories with a non-minimal matter-curvature coupling.

\section{Gravothermal Catastrophe}

From the moments of the distribution function, we can derive some macroscopic equations, such as the Navier-Stokes one for fluids, and the virial equation \cite{binney} and its cosmic version - the Layzer-Irvine equation. In fact, the latter was derived for the modified gravity model of Eq. (\ref{modelo}) in Ref. \cite{linmc} from the metric field equations. This process is preferred since it begins with the full nonlinear behaviour of the theory, in comparison with the usual one from the weak field regime to find the virial theorem. It was found that there is a deviation from the usual virial theorem given the presence of an extra potential energy term, $U_{NMC}$, in the equation which could explain the dark matter behaviour on clusters of galaxies \cite{linmc}.

In fact, heat and matter can be transferred out of a virialised and gravitationally bound system leading to a gravothermal collapse. Therefore, in the presence of a non-minimal coupling between matter and curvature, the virial equation reads $U_{\Phi}+U_{NMC}=-2 K$, where $U_{\Phi}$ is the Newtonian potential energy and $K$ is the kinetic energy. Therefore, the total energy of the system is $E=K+U_{\Phi}+U_{NMC}=-K <0$, and hence the specific heat of the system is negative, $c_V=\frac{1}{m}dE/dT$. This implies that the more energy is added to the system, the lesser the kinetic energy, and consequently, the temperature, i.e., the system becomes less strongly gravitationally bound. A relevant situation arises when heat and mass both flow radially outward in a halo with negative radial temperature gradient, and the outer halo region has a specific heat higher than the inner region which keeps contracting and increasing the temperature \cite{gravothermal1,gravothermal2}. This process is known as gravothermal catastrophe and its characteristic time is of the order of the relaxation time of the Boltzmann equation.

In what concerns the non-minimal matter-curvature coupling model, this process is not avoidable since the specific heat for a gravitationally bound system is always negative. This might, however, not be the case for other modified gravity models.

\section{Conclusions} 
\label{sec:conclusions}

In this work, we have derived the Boltzmann equation in the context of a non-minimal matter-curvature alternative gravity model. We have shown that there is an extra term which arises from geodesic deviation force. As far as the conservation laws built from the distribution function in these theories, we find that the flux density of particles is covariantly conserved, and, in opposition, the energy-momentum tensor is not. This yields a condition on the normalisation of a homogeneous distribution function. In addition, the entropy vector flux is a non-decreasing function of the spacetime coordinates as in General Relativity, which implies that the Boltzmann H-theorem is still preserved.

When considering a homogeneous and isotropic Universe, we find that the distribution function is, at least formally, equivalent to the GR counterpart, except from the fact that some quantities do not follow the GR's behaviour, as the effects of the non-minimal coupling appear on the radiation density evolution and on the matter Lagrangian choice, for instance. This implies that different cases, which were degenerated in GR, no longer yield the same solution for the distribution function.

Furthermore, in what concerns the negative specific heat in gravitationally bound systems, we find that the non-minimal coupling model does not alleviate the problem, since these theories affect the virial equation by introducing a potential energy term. However, this may not be case for other modified gravity theories.

\end{document}